\documentclass[twocolumn,showpacs,preprintnumbers,amsmath,amssymb]{revtex4}

\def\pra{{Phys.~Rev.~A}}

\def\prd{{Phys.~Rev.~D}}

\def\physscr{{Phys.~Scr}}

\usepackage{isomath}

\usepackage{upgreek} 

\usepackage{epsfig, amsmath,amsfonts, amssymb,graphicx,amsthm,color}

\usepackage{mathtools}
\usepackage{dcolumn}
\usepackage{bm}
\usepackage{rotating}
\usepackage{amssymb}
\usepackage[latin1]{inputenc}
\usepackage{graphicx}

\usepackage{epstopdf}
\begin{document}

\title{Limitations for field-enhanced atom interferometry}
\author{	 D. Comparat}
\affiliation{Universit\'e Paris-Saclay, CNRS, Laboratoire Aim\'{e} Cotton, 91405, Orsay, France}

\date{\today}

\begin{abstract}

	We discuss the possibility to enhance the sensitivity of optical interferometric devices 
	 by increasing its open area  using
	an external field gradient that act differently on the two arms of the interferometers.
	The use of combined electric and magnetic field cancel non linear terms that dephases the interferometer. This is possible using well defined (typically with $n\sim 20$ Rydberg) states, a magnetic field of few Tesla and an electric field gradient of $\sim 10$V/cm$^2$. However this allows only for  interaction times on the order of tens of $\mu $s leading a reachable accuracy of only $1$ or $2$ order of magnitude higher than standard light-pulse atom interferometers. 
	Furthermore, the control of fields and states and 3D trajectories puts severe limits to the reachable accuracy. This idea is therefore not suitable for precision measurement but might eventually be used for gravity or neutrality in antimatter studies.

\end{abstract}

\maketitle

\section{Introduction}

 The effect of an external force (gravitation
 on neutral particles, or electric and magnetic force on non neutral one) can be accurately measured using the phase acquired in the potential of a suitably built interferometer,
 as for instance demonstrated in the classic Colella, Overhauser and Werner
 experiment \cite{1975PhRvL..34.1472C}. Nowadays, most of the interferometers are  based on mechanical  gratings or optical manipulation of  internal states using 
   gratings such, as Mach-Zehnder \cite{2002NIMPB.192..129O,PhysRevLett.112.121102}, moiré \cite{2008NIMPB.266..351} or Talbot(-Lau) \cite{sala2015matter,2016PhRvA..94c3625S}  \cite{PhysRevLett.112.121102}. 
Taking the gravity measurement as a generic example,
  the RMS statistical precision $\delta g$ on
 the measurement of the $g$ value is  quite generally estimated \cite{2002NIMPB.192..129O} as
$
 \frac{\delta g}{g} = \frac{1}{C \sqrt{N_{\rm det} }}\frac{1}{\phi}  
$
 where $\phi$ is the phase difference between the paths in the interferometer.
  $C \leq 1$ is the fringe contrast and $N_{\rm det}$ the number of events detected.
 In light pulse atom interferometry, 
 the light gratings are applied at well defined times (for instance $0,\frac{T}{2},T$). 
 Therefore, the phase shift becomes independent of the atom velocity.
 For mechanical grating, with  $d$ the grating pitch, the interferometric phase shift $\phi$  is given by  $ \frac{d}{2} \frac{\phi}{2 \uppi} = \frac{1}{2} g T^2 $ 
 and the contrast of such interferometer can approach unity (the difficulty becomes now to catch all atoms by the laser pulses) \cite{2002NIMPB.192..129O}.
The
 sensitivity
 of these devices increases with the
 measured phase difference between the matter waves  $\phi = \frac{m}{\hbar} {\bm A}.{\bm g}$, which
 scales with the  enclosed interferometric  area in space-time 
 ${\bm A} =\int \Delta {\bm x} dt$.
 Increasing the area is therefore the key ingredient to improve accuracy. 
 In most of the interferometer (Ramsey-Brodé \cite{borde1989atomic} or Kasevitch-Chu \cite{kasevich1991atomic} type) the
 atomic beams are coherently
 split and later recombined using laser pulses beam
 splitters that transfer   photon momentum $\hbar k$ ($k= 2\uppi/\lambda$ is the light wavector for a wavelenght $\lambda$ that plays the role of the grating pitch $d$)
 \cite{RevModPhys.81.1051,abend2019atom}.
  In order to increase the area,
 large momentum transfer  interferometers have been demonstrated with  $ A =2 N \int t \hbar k/m dt = N T^2 \hbar k/m $ 
 using
 $N$ photons transfers from the laser beams.
 Many methods are nowadays available \cite{abend2019atom}, typically  limited to $N \sim 100$, such as Kapitza-Dirac \cite{hornberger2009theory}, Talbot-Lau  \cite{brezger2002matter,gerlich2007kapitza},
 sequential Raman pulses \cite{mcguirk2000large}, sequential two-photon
 Bragg diffraction \cite{2011PhRvL.107m0403C}, multi-photon Bragg diffraction
 \cite{muller2008atom}, Bloch oscillations \cite{2009PhRvL.102x0403M,clade2009large} or Adiabatic passage \cite{2015PhRvL.115j3001K}.

 In this letter we would like to study  another possibility that is to increase the area $A$ by using
 an external field acting differently on the two arms of the interferometer.
 A proof of principle experiment has been realized in \cite{machluf2013coherent} (see also \cite{2014EL....10563001M,margalit2018realization,amit2019t})  by the use of an external magnetic field gradient. Theoretically only the simple ideal (pure gradient) one dimensional case have been studied
 in Ref.   
 \cite{Zimmermann2017,margalit2019analysis}.
 A simple comparison can be done between  this enhanced interferometer, based on external electromagnetic accelerations $a$  where
 ${ A} \approx  \int a t^2/2 dt \approx a T^3/6 $,
 and a pure photon recoil based light interferometer, with  ${ A} = T^2 \hbar k/m $.
The
 gain exists only  when $a T \gg \hbar k/m$ so, either with $a$ very big, either thanks to a  long  interrogation time $T$.    For a typical wavelength of $\lambda = 532$ nm and for hydrogen mass atoms, this leads to $a T \gg 1 $m/s.
Because strong acceleration can be created, for instance in Rydberg states under effect of an electric field, the gain is potentially enormous. 
For instance (see detailed formula in the appendix) a Rydberg state of $n\sim 30$ and an electric field gradient of $100$V/mm$^2$ leads to a gain  $\sim 10^5$ on $\delta g/g$ compared to a standard interferometer even for very short interrogation time of $T = 100\,\mu$s. Furthermore, with lower temperature, such as the one achieved thanks to laser cooling  \cite{1988OptCo..65..419L,1993PhRvL..70.2257S}  the interrogation time  can be longer and the gain potentially much bigger.  The  enhanced interferometer
has a $T^3$  evolution compare to the $T^2$ one of standard interferometer explaining why such enhanced gradient interferometer has seemed to be so promising  \cite{stockton2011absolute,dutta2016continuous,2014EL....10563001M,margalit2018realization,Zimmermann2017,amit2019t}.
 
In our study, we will first express some experimental considerations of such gradient enhanced interferometer and show that it is not suitable for precision measurement but may  be interesting for  experiment having high temperature and low statistics such as antimatter's experiments. We then first study the 1D case and second a real 3D case but in a simple cylindrical symmetry.
  We will show that, due to Maxwell's equations, the  3D case  leads to extra terms in the phase that seems difficult to cancel. This leads to difficulties that will strongly limit the possible gain.

 \section{Experimental considerations}

 \subsection{Stability of the fields}

 For the following we assume that we can use electric and/or magnetic  external fields to act on the states.
We first stress that
 the realization a precision
 experiment using external field gradient appears  very challenging because
 the degree of control of the time and space values of the applied fields have to be on the same order of magnitude than  the relative accuracy foreseen for $\Delta g$. Because a magnetic or electric field relative spatial homogeneity, or time stability, on the $10^{-5}$ range may already seen challenging, an accuracy of  $|\Delta g/g| \approx 10^{-5}$ seems already quite hard to achieve.
 This is in stringent contrast to photon recoil based  optical interferometers
 used for high accuracy measurements
  where
 all quantities $T,k$ and $m$ can be known at very high accuracy.
But, if
 $|\Delta g/g| \approx 10^{-5}$ is well below the state of the art $|\Delta g/g| \approx 10^{-9}$ accuracy for matter waves, it will present a tremendous improve for antimatter waves, the state of the art of which being  $|\Delta g/g | \approx 100$ \cite{peters2001high,RevModPhys.81.1051,2015PhRvA..91c3629B,amole2013description,cui2018time,karcher2018improving,kritsotakis2018optimal,bongs2019taking,fu2019new,abend2019atom}).  

 \subsection{Interest for antimatter systems}

 Therefore such gradient interferometer may be used to
 study of neutral antimatter systems such as: antihydrogen $\bar H = \bar p - e^+$, positronium Ps $= e^+-e^-$, protonium Pn $ = \bar p-p$, anti-protonic helium,  their muonic counterpart, antineutrons, ..., that indeed attracts more and more attention for tests of Lorentz, CPT invariance \cite{2013AnP...525..493Y,2013arXiv1304.3721H,2014IJMPS..3060258K,2015PhRvD..92e6002K,2017arXiv171001833S}, gravity \cite{poggiani1993possible,poggiani1997measurement,Mills2002,walz2004proposal,perez2005new,2014IJMPS..3060259C,perez2015gbar}, or even for spectroscopic measurement \cite{2014IJMPS..3060257C}.
  Proposal of antimatter studies using interferometry dates back to the 1990s
 \cite{1995leap.conf..569P,1996HyInt.100..163P,phillips1997antimatter}. Most proposals are based on  imaging a first grating by a second one and then detect the particles either using a third grating or a position sensitive detector. The gratings method has
 advantages  regardless of the coherence of the source 
 \cite{chang1975space}.  That is another advantage for antimatter system   because  of the low production rate, the high temperature of the antimatter samples, meaning extended sources,
 large beam divergence and poor energy
 definition.
 
   The interrogation time $T$ is strongly related to the temperature $T_0$ of the atoms. Indeed, the particles will stay in the (laser waist or apparatus) size  $w$  typically on the cm scale only during a time $T \approx w \sqrt{m/k_B T_0}$.
 For hydrogen at 1 K this leads to time in the $100 \,\mu$s range,
 This implies the use of long lived states such  ground hyperfine, $2s$ or $nl$ Rydberg states (lifetime
 $\approx 10^{-10} n^3 (l+1/2)^2$ seconds \cite{2005JPhB...38.1765H}). For such states the typical acceleration verifies
  $ m a \sim \mu_B m_F \nabla B$ in a magnetic field $B$, with $ \mu_B $ is the Bohr magneton and $m_F$ the total magnetic quantum number. Thus,
 the  $a T \gg 1 $m/s condition   indicates that 
 the
 enhanced interferometer 
 has increased performances compared to 
 light interferometer, as the one proposed in
 \cite{PhysRevLett.112.121102,2002NIMPB.192..129O}, only when  (for hydrogen mass)
 $m_F \nabla B \gg 1 $ T/m magnetic field gradient or $ n^2 \nabla F \approx 10^6$ V/m$^2$ for the electric field. Both conditions seems easy to be achieved and it is therefore worth to investigate in more detail the setup. 
 
  \subsection{Simple example:  1D picture, pure gradient  acting in a two arms interferometer}
 
 Even if we see that such poorly accurate interferometer will interest mostly only the antimatter community, our study of a gradient enhanced area interferometer will be more general.
We focus our general  discussion on simple two arm interferometer for sake of simplicity but most of the discussion would be relevant for multiple arms or  multiples gratings.
 Different forces acting on the  two arms of the interferometers  requires at least two different internal states $|1\rangle$ and $|2\rangle$ on which the external field produce  different accelerations ${\bm a}_1$ and ${\bm a}_2$. 
The force acting differentially on both states will create two well spatially separated arms  \cite{2015PhRvA..91c3629B,2016PhyS...91e3006B} with spatial and internal states entangled. 
This has to be compare to  Talbot-Lau setup or  classical moir\'e deflectometer, where this entanglement does not exist, and so where the final  measurement has to be spatially resolved. In our case the,  final measurement at the output of the interferometer can be performed by a simple internal state measurement. This allow to work with hotter beams and smaller spatial displacement compared to single internal state interferometer \cite{PhysRevLett.112.121102}.

  The simplest example uses two classical trajectory paths (the internal state can change along the path) through the
 interferometer with phase evolutions $\phi_1$ and $\phi_2$ the fringe phase shift will be given by $ \Delta \phi=\phi_1-\phi_2$.
 As studied in Ref. \cite{Zimmermann2017,amit2019t},
 the simplest ideal case uses a 1D picture with a pure field gradient
 $E$ (electric or magnetic with $E= \| {\bm E} \| = E_0 + E' (z-z0)$), a linear field dependence on the potential energy ($E_{\rm p}(E) = E_{\rm p}(E_0) + E'_{\rm p}(E_0) (E-E_0)$ for a given state). This creates uniform,  non spatial neither time dependent acceleration $\tilde a_1$ and $\tilde a_2$ with $\tilde a_i = E' {E'_{\rm pot}}_i (E_0)/m$. So including the gravity,
the two internal states $|1\rangle$ and $|2\rangle$ have different accelerations  $a_1=g+\tilde a_1$ and  $a_2=g+\tilde a_2$. 

\begin{figure}
	\centering
	\includegraphics[width=\linewidth]{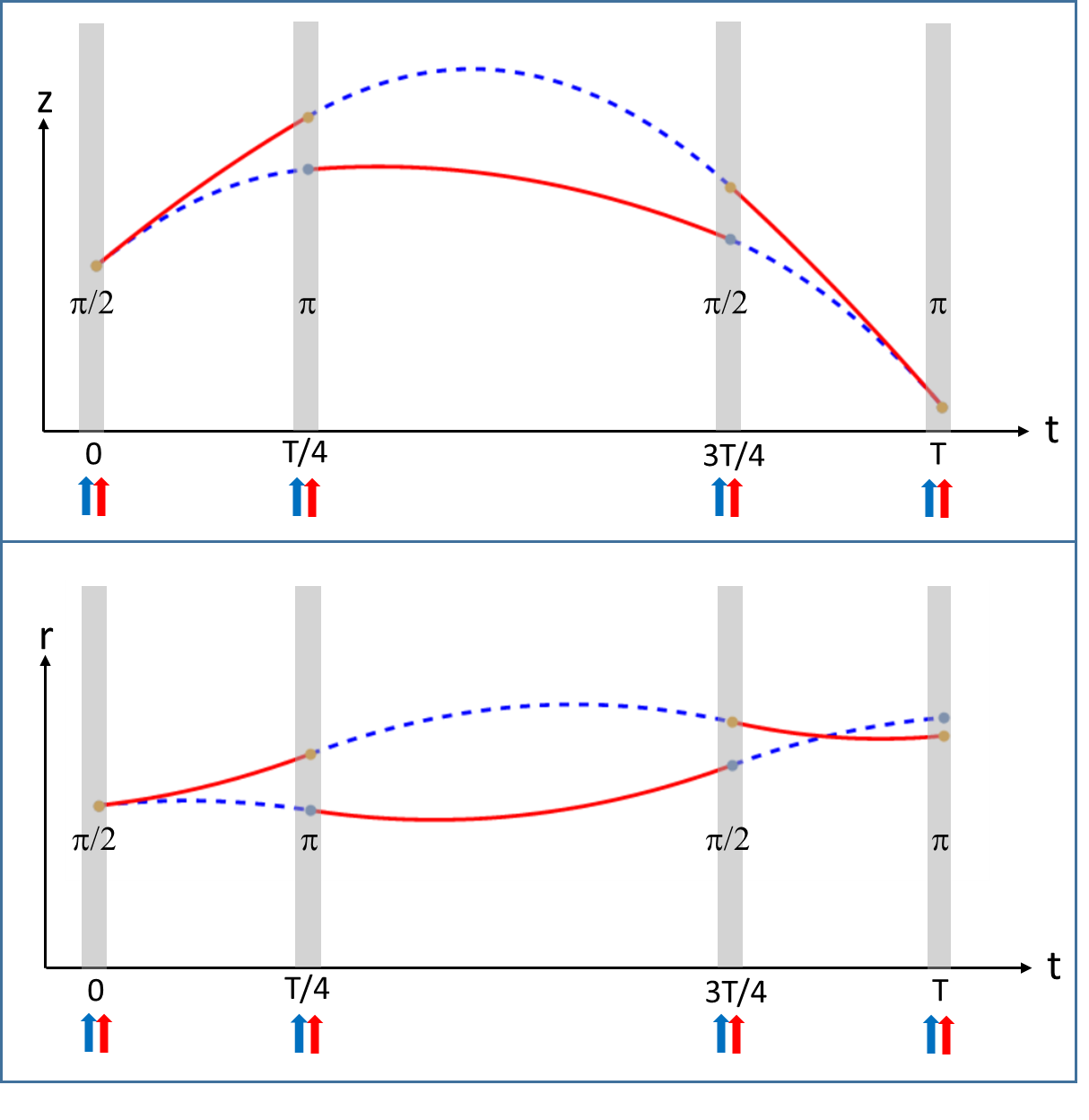}
	\caption{Space-time diagram of the gradient enhanced interferometer with 
		four  (short and co-propagating Raman) laser pulses at time  $0,T/4,3T/4,T$. The two internal states are shown receptively in solid-red and dashed-blue curves.
		If seen as a pure 1D interferometer a pure gradient interferometer can be closed in $z$ coordinate (upper panel). However when looking to 3D, or with non linear terms,  the situation is more complex as shown for the radial coordinate $r$ (lower panel).}
	\label{fig:interferometer}
\end{figure}

To close in phase space the interferometer several timings and accelerations are possible.  Choosing one solution, or the other, does not change the conclusions we are going to derive. One possible solution is given in Ref.   \cite{mcdonald2014faster} where
 (forgeting the gravity) the acceleration of the upper part is $\tilde a, - \tilde  a, 0, 0$ and $0,0, \tilde a,- \tilde a$ for the lower path, leading to
$   \Delta \phi = -\frac{m}{\hbar} g a T^3 /32$ where $T$ is the total time spend in the interferometer. This solution uses 
three different accelerations, so in principle three different internal states. Therefore,
we will use here the simpler solution proposed in Ref. \cite{Zimmermann2017},
and recently realized
in Ref. \cite{amit2019t} (with $\pi$ pulse being replaced  by field reversals), because
it  only requires two accelerations $\tilde a_1$ and $  \tilde  a_2$. Following \cite{Zimmermann2017} we will thus
create (by a $\pi/2$ pulse) the superposition at time $0$ then changing states ($\pi$ pulse) at time $T/4,3T/4$ and recombining ($\pi/2$ pulse) at $T$, as illustrated in Figure \ref{fig:interferometer} upper panel.

This choice closes the interferometer both in position and velocity meaning that
the classical path $\Gamma$ linking  the initial to the final point,
through  Newton's classical trajectories equation ${ m {\ddot {x}}_{i} + {\frac {\partial E_{\rm p}}{\partial x_{i}}} = 0}$, ends at the same phase space position for both arms \cite{Zimmermann2017}.
The phase imprinted by the lasers is $\phi_L = \phi_L(0)-2 \phi_L(T/4) +2 \phi_L(3T/4) - \phi_L(T)$  \cite{Zimmermann2017,zimmermann2019representation}.   Because being zero if assuming no phase jump (coherent laser)
we will  neglect it in the following. Thus,
 in this simple case the calculation of the phase evolution (see Eq. (\ref{phase_calc}) detailed later) is straightforward and leads to
 $\Delta \phi = \frac{m}{\hbar} (a_1^2 -a_2^2) T^3/64 = 
 -g  \frac{m}{\hbar} (\tilde a_1 -\tilde a_2)  T^3 /32
 +  \frac{m}{\hbar} (\tilde a_1^2 -\tilde a_2^2)  T^3/64
 $.
 This formula confirms that, in addition to the desired gravity dependent phase, another global phase, insensitive to gravity, arises. Because we want  the gravity term big, we want  the accelerations to be big and thus this second phase can also be quite big.  Using  $\tilde a_1 \approx -\tilde a_2$  limits the value of this extra phase. However without specific tricks it will be difficult to experimentally ensure perfectly this  equality because 
$ \tilde a_1 $ and $\tilde a_2$ 
are sensitive to field fluctuations. Consequently, the field fluctuations will either reduce the contrast or even blur (if this extra phase is bigger that $2\pi$) the interferometer.

 \subsection{Real 3D case}

 In addition to this important difficulty, another problem exists. Indeed,
 the one dimensional (1D) picture is clearly not adequate because  Maxwell's equation (or Gauss's law) implies that an electric or magnetic field gradient   acts on a multidimentional space. Therefore the electromagnetic force created by an external field gradient is necessary acting on, at least, two directions. This is important because
  to keep a good contrast, the trajectories, and so the area, have to be the same for all initial positions and velocities. This can be achieved, for instance, if the particles are submit to a  force that does dot depends on positions neither on velocities. As we will see, when combined with the non linearity of the Stark or Zeeman effect, this  puts strong limits on the states and fields that could be chosen. 

      Maxwell's equations for static electric and magnetic fields in vacuum have the exact same form so we can treat both field in the same way. 
      We shall thus first, for simplicity, use an electric field and a simple cylindrical symmetry along the vertical $z$ axis.
      From the sole voltage along the axis $V(z)$,  Maxwell's equations implies a single solution for 
      the 3D potential (in cylindrical coordinates) $V(\rho,z)$.
      Indeed, using series notations
      $V(z)= V_0 - E_0 z - E' z^2/2 - E'' z^3/6 - \cdots$ we find \cite{Orloff2008}:
     $$V(\rho,z) = V_0-E_0 z-E' z^2/2-E'' z^3/6+1/4 r^2 (E'+E'' z) + \cdot$$ 
      The gradient gives the electric field $\bm E$.
      Taking the series (in $E_0$ or similarly in $r,z$), up to the third order, for  the norm $E=\| \bm E \|$ leads to
 $$
      E = E_0 + E' z + \frac{E'' }{2} z^2 +  \frac{E'^2 - 2 E_0 E'' }{8 E_0} r^2   - E' \frac{E'^2-2 E_0  E''}{8 E_0^2} z r^2 
  $$
  
  Similarly the potential energy can be written is series: $$E_{\rm p} (E) =E_{\rm p}(E_0) + E'_{\rm p}(E-E_0)  + \frac{1}{2} E''_{\rm p}(E-E_0)^2 +  \frac{1}{6} E'''_{\rm p}(E-E_0)^3  $$
   Where we have noted $E'_{\rm p}$  for $\frac{{\rm d}  E_{\rm p}}{{\rm d} E}(E_0)$. Finally,
  up to the third order, the Lagrangian becomes:
    \begin{eqnarray}
    L&=&  \frac {p_z^2}{2m} + \frac {p_r^2}{2m} -
  E_{\rm p}(E_0) - E'_{\rm p} E' z   - m g z \label{Lagrangian} \\
    & & -    \frac{E'_{\rm p} E''  + E''_{\rm p} E'^2 }{2} z^2  -  E'_{\rm p}  \frac{E'^2 - 2 E_0 E'' }{8 E_0}  r^2   
    \nonumber \\
    & & - E'  \frac{E'^2-2 E_0  E''}{8 E_0^2} (  E''_{\rm p} E_0 - E'_{\rm p} )  r^2 z  \nonumber \\
    & & -  \frac{E'''_{\rm p}E'^3+ 3 E''_{\rm p} E' E'' }{6} z^3
    \nonumber
    \end{eqnarray}

 To calculate the phase evolution, we will separate this full lagrangian  ${ L({\bm {x}},{\dot {\bm {x}}},t) = \frac{1}{2} m  {\dot {\bm {x}}}^2  - E_{\rm p}({\bm {x}},t)}$ between a quadratic
    lagrangien $L_0$ 
    containing   only homogeneous acceleration and its gradient (so the first 2 lines of Eq. (\ref{Lagrangian})), and  a perturbative Lagrangian $L_1$  
    containing only the third order terms ($r^2 z$ and $z^3$ ones).
 As recalled  in the Appendix, see Eqs. (\ref{phase}) and (\ref{delta_phase}), if
 $L_1$ is  a perturbation, 
 the evaluation of the total phase under the $L$ lagrangian, can  be evaluated using  only  the  classical unperturbed path $\Gamma_0$ derived with the sole $L_0$ lagrangian:
 \begin{equation}
 \phi = \frac{1}{\hbar} \int_{\Gamma_0} L({\bm x}, \dot{\bm x},t) {\rm d} t  {\rm d}t 
 \label{phase_calc}
\end{equation}
 We use this formula all along in this article 
and,
because of its perturbative nature, we will also use only formulas up to the third order in time.

    The first line of Eq. (\ref{Lagrangian}) correspond to the 1D case due to the fields gradient $E'$ and the linear dependence of the potential energy $E'_p$.  However in 3D, we  see extra terms appearing.
    One of the most   important is
    $  E'_{\rm p}  \frac{E'^2  }{8 E_0}  r^2  $ that occurs  even in a pure  gradient ($E'$) along $z$ and a  linear potential energy ($E'_{\rm p} $) studied previously. This illustrates already why the  3D picture is required even to study the simplest "1D-case". Even if the gradient of the field is perfectly homogeneous (and the dependence of the potential energy is  linear), some non linear  terms are present in the lagrangian. 
    Therefore even in this ideal case the problem can not be treated as a 1D problem as done in \cite{Zimmermann2017,amit2019t}.      
    
   More generally, the presence of extra terms will produce (very large) extra phases (see for instance in Figure \ref{fig:interferometer} lower panel)
 that we thus need to cancel, or control, as much as possible.

  \section{Electric and magnetic field cancellation}
     
         \subsection{Cancellation using pure field}
   
     Compared to the ideal case, some quadratic and even non quadratic terms arise. 
      A simple solution would be to cancel these extra terms by appropriate choice of fields or states.
    For instance an appropriate choice of the external field with
      $ E'' = \frac{E'^2 }{ 2 E_0}  $ would cancel the $r^2$ term (and also the  $r^2 z $ term); and
     an appropriate choice  of a  state with a potential energy that fits $E''_{\rm p}  = - \frac{E'_{\rm p}  }{ 2 E_0}  $ would cancel the $z^2$ term. We note that the $z^3$ term could, in principle also be canceled by choosing
      $E'''_{\rm p} = -  \frac{3 E''_{\rm p}}{2 E_0}$;
        but because the state, and so its potential energy  $E_{\rm p}(E)$, is already partially imposed by the previous equality we may have not enough degree of freedom for the choice.
        
But, even canceling only the second order terms seems difficult because if this requires  to have the equation  $E''_{\rm p}  = - \frac{E'_{\rm p}  }{ 2 E_0}  $ verified only locally, that is at field $E_0$; we  see that, if we want to solved it for all possible $E_0$, it becomes equation $E''_{\rm p}  = - \frac{E'_{\rm p}  }{ 2 E}  $ that is  solved in   $E'_{\rm p}  ( E) \propto \sqrt{E} $. Thus, to cancel the second order terms the variation of the potential energy should, at least locally at field $E_0$, have a kind of a square-root dependence.
Unfortunately, Eq. (\ref{eq_E}), for (anti-)hydrogen atoms in Rydberg state
 indicates that
 the Stark effect is quite linear.
We have indeed
 checked that
the cancellation
 is impossible for all (anti-)hydrogen states because $E''_{\rm p} <
 -\frac{E'_{\rm p}  }{2 E_0}$, for any electric fields below the  ionization threshold ($1/9 n^4$ in atomic units  for a level with a principal quantum number $n$). 
 Similar behavior exists for pure magnetic Zeeman effect for low lying states.
  We have also  checked that it is impossible to get $E''_{\rm p}  = - \frac{E'_{\rm p}  }{ 2 E_0}  $ for the (anti-)hydrogen ground state even if taken into account the hyperfine structure using the Breit-Rabi formula. 
 
 It is however interesting to note that the Breit-Rabi formula for nuclear spin value $I>1/2$  allows the equality $E''_{\rm p}  = - \frac{E'_{\rm p}  }{ 2 E_0}  $ for levels sub-Zeeman levels $m_F<0$. This in addition occurs for the interesting case of 
  $\tilde a_1 \approx -\tilde a_2$  for the two $|F=I\pm1/2,m_F\rangle$ hyperfine states.
  For instance for
   $^{87}$Rb it arises at $B=0.0440665\,$Tesla for both  $|m_F=- 1\rangle$ states and for $^{85}$Rb  
it arises at  $B=0.0283573\,$Tesla for  both $m_F  = - 2$ and
at $B=0.0124467\,$Tesla for both $m_F  = - 1$~\footnote{Daniel A. Steck,  available online at http://steck.us/alkalidata}. 
It is beyond the scope of this article to cover all atomic cases including fine, hyperfine or diamagnetic terms. But,
the first conclusion concerning the cancellation is that
 (anti-)hydrogen atom is peculiar because of its almost linear Stark (and Zeeman) effect.
The simplest possibility to cancel the non linear terms is thus to combine electric and a magnetic field to produce locally  a square-root dependence of the potential energy curve with the fields. We are going to study this case that will also include the fact that
 for Rydberg states the diamagnetic Zeeman effect can play a substantial role
\cite{lisitsa1987new,friedrich1989hydrogen,pinard1990atoms,gallagher1994,Bartsch2006}.

\subsection{Cancellation using electric and magnetic field}

Combining electric and magnetic field
 modifies strongly the curvature of the energy levels and provides some level crossings. This might thus lead to the cancellation we are looking for. 
For this study we simply use 
 the (second order) potential energy formula,  Eq. (\ref{eq_EB}),  valid for arbitrary $\bm B$ and $\bm E$ fields. The eigenstates are noted $|n,m_1,m_2\rangle$ 
 with $m_1$,$m_2$ quantum numbers spanning $-(n-1)/2,-(n-3)/2, . . . ,(n-1)/2$ .
 
Several fields geometry  are possible. We have looked to many of them and found similar results, therefore we illustrate the result using only the simplest case of a constant and uniform magnetic field along the $z$ axis, in addition to the already studied cylindrical symmetric electric field. 
 This solution is appealing because such 1-5 Telsa field is naturally present in
most of the antihydrogen experiments.

For these fields, the potential energy is  calculated using 
Eq. (\ref{eq_EB}). We then analytically expand it in series to evaluate the $E'_{\rm p}, E''_{\rm p}$  terms.
This leads to the choices (in atomic units)
\begin{eqnarray*}
E_0 &= & 0 \\
\frac{E''}{E'^2} &=& \frac{ n^3 (-19 + 12 (m_1^2 + m_1 m_2 + m_2^2) - 17 n^2)}{
	12 ( m_2 -m_1 ) } \\
B &=& -\frac{12 (m_1 + m_2)}{n^2 (21 - 20 m_1 m_2 + 15 n^2)}
\end{eqnarray*}
in order to cancel the quadratic terms in the lagrangian. This choices create a linear potential energy 
$E_p  = -(3/2) (m_1-m_2) n E' z$
(up to the second order terms)  for a given $|n,m_1,m_2\rangle$  state. The third order terms are \textit{a priori} not cancelled.

Cancellation of the second order non linear terms (in $r^2$ and $z^2$) imposes the value for the magnetic and electric field for a given $|n,m_1,m_2\rangle$  state. But
 we have (at least) two states in the interferometer. 
 Thus we have to choose compatible $|n,m_1,m_2\rangle$ (for state 1 with acceleration $\tilde {\bm a}_1$) and $|n',m'_1,m'_2\rangle$ (for state 2 with acceleration $\tilde {\bm a}_2$) levels that gives similar (exact equality was found to be impossible) magnetic and electric fields. So, for instance, with 
 $ B \approx -\frac{12 (m_1 + m_2)}{n^2 (21 - 20 m_1 m_2 + 15 n^2)} \approx -\frac{12 (m'_1 + m'_2)}{n'^2 (21 - 20 m'_1 m'_2 + 15 n'^2)}$
 and, if possible, that do not create too high third order terms.

Several choices of pair of states are possible. We found very similar results with many choices and so, we simply mention 3 of them to express possible and typical values for the interferometric phases:

\begin{itemize}
	\item
	$n = 20,  m_1 = 13/2,  m_2 = -19/2 $
	and
	$n' = 21, m_1' = -1 , m_2' = -2$  requires similar magnetic fields  $B_0 \approx 2.91\,$T and  $E'^2/E''' \approx 200 $V/cm.
	\item  
	$n = 19,   m_1 = -7,  m_2 = 3,   $  and
	$n' = 20, m_1' = -17/2, m_2' = 7/2  $  with $B_0 \approx 5.33\,$T and  $E'^2/E''' \approx -160 $V/cm.
	\item 
	$n = 18,  m_1 = 11/2,  m_2 = -17/2  $
	and $ n' = 37 , m_1' = -18, m_2' = -15 $
	 with $B_0 \approx 4.49\,$T and  $E'^2/E''' \approx 300 $V/cm
\end{itemize}
 
Because each chosen pair of states leads to results within the same order of magnitude of the final phases, we give here the results using only the third (last) choice.

  In order to quickly estimate the contrast of the interferometer we calculate the  phase evolution for  16 particles with two different initial  position in $z$, two different initial  position in $r$, two different initial  velocities along $z$ and two different initial  velocities along $r$. We compare the phase with the one arising from a particle starting at the center and with zero initial velocity. We finally  averages the absolute values of all phases.
We use 
 a  typical (axial $z$ and radial $r$)  distance of 1 mm and axial and radial thermal velocities corresponding at  an initial temperature of $ T_0=0.01 $ Kelvin ($\sqrt{k_B T_0/m} \approx 10\,$m/s for the (anti-)hydrogen mass). 
   This temperature is chosen because it is reachable with laser cooling methods and 
   higher temperature starts to create too big dephasing and lower one does not really help because the dephasing is not anymore dominated by the velocity but by other effects such as the third order terms in the Lagrangian.

  The cubic dependence of the phase with the evolution time favors long evolution times. However in order to keep dephasing (due to not perfect cancellation of second and third order terms in the lagrangien for both states simultaneously) smaller than $2\pi$ we restrict to
an evolution time (between light pulses) of  $T/4 = 100\,\mu$s. We choose an  electric field gradient of 
$E' = 10\,$V/cm$^2$. It cannot be much larger, to avoid too big dephasing terms and it cannot be much smaller because it is what creates the opening of the area of the interferometer and so the enhancement effect we are looking for.
For our gradient interferometer we find  a total (global) phase of
$\phi =
472808$ ($\phi \propto E'^2 m^{-1} T^3$).
This indicates that, as previously stated, a field stability and geometric homogeneity on the order of $10^{-5}$ is probably require in order to avoid a complete blurring of the fringes (meaning a 
fluctuation of this phase value of less than $2\pi$).
Unfortunately, the interesting term linked to gravity $2 \frac{m}{\hbar} g (a_2-a_1) T^3$ is only 
$85.85$ (this term is proportional to $E' m^{0} T^3 $).
Therefore the gain (factor $\sim 30$) compare  to a simple   Kasevich-Chu phase of $k g T^2 = 3.08$ (calculated with a quite arbitrary choice of $k = (2\pi)/(200 nm) $) can be seen as marginal compare to the optimistic value expected at the beginning of this article.
Finally, the
 (error) phase 
due to second and third order terms is indeed small ($  0.3$) and will probably not create a too bad contrast.

\section{Conclusion}

We have  shown that using  a pure gradient of electric or magnetic field in order to produce strong forces able to increase the area of an interferometer, that seems attractive when looking  on a 1D picture, turns out to be quite complicated in a real 3D picture. Despite the fields stability issues the other limitations are:
first, a pure gradient along one axis  is not possible due to zeros of the field divergence in Maxwell's equation; second, the extra terms produced are difficult to cancel using pure fields for (anti-)hydrogen atoms. We nevertheless note that for ground state alkali atoms the Breit Rabi formula allows this cancellation for specific magnetic field values, that might be interesting to be studied.

Combining electric and magnetic field  helps but has many drawbacks (in addition to the technical difficulty!) because the cancellation is not perfect implying  to choose  given fields values and geometries as well as choosing proper  states. 
Furthermore, the gain we found was quite marginal, at best a factor 100 compared to a simple  Kasevich-Chu's interferometer (with $N=1$).  We have studied only some particular fields geometry and pulse sequences but we doubt that other geometry, time sequences or even time dependant fields, would lead to drastically  better results.

In conclusion, we do not see any strong advantages for precision measurement to use such enhanced interferometer compare to standard ones. We therefore did not discuss in detail the practical implementation of such interferometer. It is nevertheless worth mentioning that  matter-wave interferometry with 
hydrogen
atoms in  Rydberg states has been already demonstrated
in \cite{heupel2002hydrogen} (by coupling $2s$ and $15p$ levels) and that interferometer using only high Rydberg states have also recently been demonstrated
\cite{palmer2019electric,palmer2019matter}. We simply mention that one advantage of such a scheme, using Rydberg states, is than RF or microwave pulses can be used that are easy to implement and can address more velocity classes (due to the reduced Doppler effect) that Raman laser pulses.
Therefore it still might be of interest, for antimatter experiment, to consider such schemes.

We finally mention one possibility to improve the result. We have used a perturbation methods, for the fields and for the potential energy, to study the problem, but it is possible that  strong non linear terms can be more efficient. We can  think of a so strong second order terms that it would lead to trapping of the particles and, as  after one oscillation period in a pure harmonic trap, the interferometer can even be closed. This might leads to interesting interferometric measurement that might deserve more studies.

\begin{acknowledgments}

We thanks J. Robert, P. Cheinet and P. Yzombard for fruitful discussions.

\end{acknowledgments}

\appendix

\section{Stark Effect}

In an electric field $F$, neglecting the fine structure effects that are small for Rydberg states,
the
energy of the states  $E = E_{n,n_1,n_2,m}$ can be accurately calculated from  the fourth order
expansion of the hydrogen Stark \cite{1978PhRvA..18.1853S,lisitsa1987new,PhysRevA.47.1209,PhysRevA.87.063423} states, in atomic units:
\begin{eqnarray}
E &=& -	\frac{1}{2 n^2} + \frac{3 n}{2} k F - \label{eq_E}  \\
& & 
\frac{n^4}{16}  (17 n^2 - 3 k)^2 - 9 m^2 + 19 ) F^2 + \nonumber \\
& &
\frac{3 n^7}{32} k (23 n^2 - k^2 + 11 m^2 + 39) F^3 - \nonumber \\
& &
\frac{n^{10}}{1024} \big( 5487 n^4 + 35182 n^2  - 1134 m^2 k^2 + \nonumber \\
& &
1806 n^2 k^2 - 3402 n^2 m^2 + 147 k^4 - \nonumber	\\
& & 549 m^4 + 5754 k^2  - 8622 m^2 + 16211 \big) F^4 \nonumber
\end{eqnarray}
where $n = n_1 + n_2 + |m| + 1$ and $k=n_1-n_2 = 2n_1 -n -|m|$.

Back to SI units ($4.36 \times 10^{-18}{\rm J}$ for the energy and $5.14 \times 10^{11}{\rm V/m}$ for the field),
the acceleration that can be created in an electric field  $F$ is
 on the order of  $m a  \sim 4.36 \times 10^{-18}{\rm J}  \frac{3}{2} n^2 \frac{\nabla F}{5.14 \times 10^{11}{\rm V/m}} $




\section{Stark-Zeeman Effect}

The  energy levels of an hydrogen atom in
electric and magnetic fields with arbitraty orientation has been studied \cite{PhysRevA.57.1149} and analytical formula up to the 
second-order in the fields have been calculated 
\cite{solovev1983second}. The states are labeled $|n n' n'' \rangle$; they correspond when $E=0$ to the $|n n_1 n_2 m \rangle$ Stark states with $n'=(m+n_2-n_1)/2$ and $n''=(m-n_2+n_1)/2$ \cite{demkov1970energy}.
In atomic units ($f=F/(5.14 \times 10^9$ V/cm , $\gamma = B/(2.35\times 10^5$ T)) we have:
\begin{eqnarray}
E_p &=& -	\frac{1}{2 n^2} + |{\bm \omega}_1| n' +  |{\bm \omega}_2| n'' \label{eq_EB}  \\
& & 
- \frac{n^4 f^2}{16} \{
17 n^2 + 19 -12[ n'^2 + n' n'' \cos(\alpha_1+\alpha_2)+n''^2]
\}  \nonumber \\
& &
+ \frac{n^2 \gamma^2}{48} \{
7 n^2 + 5 +4 n' n'' \sin\alpha_1 \sin\alpha_2  \nonumber \\
& &
+(n^2-1)(\cos^2\alpha_1 + \cos^2\alpha_2)  \nonumber \\
& & -12(n'^2 \cos^2\alpha_1 - n' n'' \cos\alpha_1 \cos\alpha_2  \nonumber \\
& &
+n''^2 \cos^2 \alpha_2)]
\nonumber 
\end{eqnarray}
where ${\bm \omega}_{1,2} = \frac{1}{2} ({\bm \gamma} \mp 3 n {\bm f})$ and $\alpha_1$ and $\alpha_2$ are the angles between the magnetic field axis and the vectors ${\bm \omega}_{1}$ and ${\bm \omega}_{2}$ respectively.

\section{phase evolution in interferometers}

\label{phase_interferometer}

We recall here the basics of the calculation of the evolution of the atomic wavepacket in our simple two path atom interferometer. We will deals with non
relativistic atom interferometry (see \cite{borde2014atom} for a more general case).
Several methods exists to study the evolution:  using a plane wave or a gaussian wavepacket decomposition  \cite{borde2001theoretical,antoine2003quantum}, sometimes linked with a path integrals formulation \cite{kasevich1991atomic,RevModPhys.81.1051}, or using the density matrix \cite{dubetsky2006atom} or Wigner function evolution \cite{2014NJPh...16l3012R,kleinert2015representation,roura2017circumventing,chaneliere2018phase,zimmermann2019representation}). Obviously, all methods leads to the same final results
but the choice is made depending on the context.

The most important results concern the case of the (laser free) evolution under an hamiltonian (or a lagrangian) containing a potential that is at most quadratic in the space coordinates. That is when
 each internal state  $|1\rangle,|2\rangle,\cdots$ evolves under its own hamiltonian $H_1,H_2,\cdots$
that contains only  homogeneous acceleration and its gradient. This is for instance  the case for the first 2 lines of Eq. (\ref{Lagrangian}).
 Under such circumstances, that fortunately are the most usual ones,  the
quantum phase-space (Wigner)  distribution
 evolves under the same (Liouville's) equation than the classical phase-space distribution \cite{case2008wigner,chaneliere2018phase}.  The fact that the evolution 
 is given by the classical evolution 
  is also
 directly visible using
 the evolution operator between the time $t_i$ and $t_f$ that, in the 2 levels case 
 $\hat U(t_i,t_f) = \begin{pmatrix}
 U_{11} (t_i,t_f)	 & 0 \\ 
 0  & 	U_{22} (t_i,t_f)
 \end{pmatrix} $,
is given in the position representation (here written in 1D to simplify the notations)  by
$\langle z_f | \hat U_{11} (t_i,t_f) | z_i \rangle= \sqrt{\frac{m}{2 i \pi \hbar (t_f-t_i) }  } e^{i S_{\rm cl}(z_f,t_f ; z_i t_i)/\hbar}$ where $S_{\rm cl}$ is the classical action evolution of a particle in state $|1\rangle$, starting at position $z_i$ at time $t_i$  under the forces created by $H_1$ and arriving at time $t_f$ at position $z_f$.
In other words
the phase evolution of the wave function is given by  the (semi-classical limit of the) Feynman's path-integral formulation in terms of the
Lagrangian by 
\begin{equation}
\phi = \frac{S_{\rm cl}}{\hbar} = \frac{1}{\hbar} \int_{\Gamma } L({\bm x}, \dot{\bm x},t) {\rm d} t \label{phase}
\end{equation}
and $\Gamma$
is the classical path from the initial to the final point.
This was first proposed by Kennard and Van Vleck \cite{kennard1927quantenmechanik,van1928correspondence}
and  shown by Morette \cite{morette1951definition}
to be exact in the quadratic case. 
 
In our case, in addition, we deal with higher order terms, such as
the third order ones ($r^2 z$ and $z^3$ ones). The theory is more complex and generally no more  linked to the classical world with for instance negative values for the Wigner function
\cite{case2008wigner,chaneliere2018phase}. Fortunately, if
 the extra terms are considered as a perturbation  $L_1$ on the Lagrangian $L = L_0 +  L_1$, the phase shift $\delta \phi$ introduced into the final wavefunction, by the perturbation $L_1$, is given simply (to the first order) by the integral of the perturbation along the classical unperturbed path $\Gamma_0$ \cite{storey1994feynman,bongs2006high,bertoldi2019phase}: 
 \begin{equation}
  \delta \phi = \frac{1}{\hbar} \int_{\Gamma_0} L_1 {\rm d}t
  \label{delta_phase}
 \end{equation}
This allows to calculate the phases created by the accelerations.

We should also deal with the $\pi$ or $\pi/2$ light pulses.
 The general case of the interaction with lights can be complicated because
correlations may appear between internal and external
variables invalidating the Bloch-equation or the simple semi-classical approaches
\cite{case2008wigner,chaneliere2018phase}.
However, because  it is not the main focus of our article to deal with these issues, we will restrict ourselves to ideal case of quasi-instantaneous light pulse and with
no momentum transfer created by the  pulses. This can be realized  for instance thanks to the use of co-propagating laser Raman beams.
Therefore, for
$\pi$ pulse at time $t_i$ we have
$\hat U_\pi (t_i) = \begin{pmatrix}
0 	 &  -i e^{ i \phi_L(t_i)} \\ 
-i e^{- i \phi_L(t_i)}  & 0 
\end{pmatrix} $. Similarly, the evolution under a $\pi/2$ pulse is given by
$\hat U_{\pi/2} (t_i) = \frac{1}{\sqrt{2}} \begin{pmatrix}
1	 &  -i e^{ i \phi_L(t_i)} \\ 
-i e^{- i \phi_L(t_i)}  & 1 
\end{pmatrix} $. In these formula   $\phi_L(t_i)$ is the laser phase at time $t_i$ (or the phase difference $\phi_L(t_i) = \phi_{L_1}(t_i) - \phi_{L_2}(t_i)$  in the case of two Raman laser 
beams).

In conclusion, by multiplying the matrix evolution we can calculate the state evolution.
In our case of simple  (upper  $u$ and lower  $l$) paths  with the 2 internal states;
if  starting, (for instance) with atoms in state $|1\rangle $, we find that the probability to observe atoms in the state $|1\rangle $  just after the last $\pi/2$ pulse is
$$P = \frac{1}{2}(1- C \cos(\phi_u - \phi_l +\phi_L) )$$ where
$\phi_u $, respectively $\phi_l $, is the phase acquires by the particles in the upper,  respectively lower branch (with possible internal state change during the motion).
$\phi_L$
comes from the phases imprinted by the lasers.
For a single atom the contrast $C=1$, but
obviously when summed over all initial atomic phase space densities, the dephasing and the incoherent sums of the probabilities  lead to a reduction of the contrast. 
A simple  case is when the 
initial wavepacket  ($|  \psi_0 \rangle$ in the pure case), or the initial  phase-space distribution (in the statistical ensemble  case), is   gaussian  and the evolution is quadratic. In such case, as shown before,  the position-momentum mean and (co-)variance  evolves in very simple an analytical manner given by the classical evolution (in the so-called  
ABCD $\xi$ theorem) \cite{borde2001theoretical,antoine2003quantum,2014NJPh...16l3012R,kleinert2015representation,roura2017circumventing,zimmermann2019representation})  directly giving the final contrast (using obvious notations)
$C e^{i \phi} = \langle \psi_0 |\hat U_u^\dag \hat U_l |  \psi_0 \rangle = 
\langle \psi_u  |  \psi_l \rangle = \int \psi_u^* (z)  \psi_l (z) {\rm d} z$. However our case is more complex with non quadratic terms in the hamiltonian. No analytical formulas exists, and we therefore simply estimate the lost of contrast, or dephasing, by calculating the phases for several different initial positon-momentum  states.



\begin{thebibliography}{10}
	
	\bibitem{1975PhRvL..34.1472C}
	R.~{Colella}, A.~W. {Overhauser}, and S.~A. {Werner}.
	\newblock {Observation of Gravitationally Induced Quantum Interference}.
	\newblock {\em Physical Review Letters}, 34:1472--1474, June 1975.
	
	\bibitem{2002NIMPB.192..129O}
	M.~K. {Oberthaler}.
	\newblock {Anti-matter wave interferometry with positronium}.
	\newblock {\em Nuclear Instruments and Methods in Physics Research B},
	192:129--134, May 2002.
	
	\bibitem{PhysRevLett.112.121102}
	Paul Hamilton, Andrey Zhmoginov, Francis Robicheaux, Joel Fajans, Jonathan~S.
	Wurtele, and Holger M\"uller.
	\newblock Antimatter interferometry for gravity measurements.
	\newblock {\em Phys. Rev. Lett.}, 112:121102, Mar 2014.
	
	\bibitem{2008NIMPB.266..351}
	A~Kellerbauer, M~Amoretti, A.~{}S. Belov, G~Bonomi, I~Boscolo, R.~{}S. Brusa,
	M~B\"{u}chner, V.~{}M. Byakov, L~Cabaret, C~Canali, C~Carraro, F~Castelli,
	S~Cialdi, M~de~Combarieu, D~Comparat, G~Consolati, N~Djourelov, M~Doser,
	G~Drobychev, A~Dupasquier, G~Ferrari, P~Forget, L~Formaro, A~Gervasini,
	M.~{}G. Giammarchi, S.~{}N. Gninenko, G~Gribakin, S.~{}D. Hogan, M~Jacquey,
	V~Lagomarsino, G~Manuzio, S~Mariazzi, V.~{}A. Matveev, J.~{}O. Meier,
	F~Merkt, P~Nedelec, M.~{}K. Oberthaler, P~Pari, M~Prevedelli, F~Quasso,
	A~Rotondi, D~Sillou, S.~{}V. Stepanov, H.~{}H. Stroke, G~Testera, G.~{}M.
	Tino, G~Tr\'{e}nec, A~Vairo, J~Vigu\'{e}, H~Walters, U~Warring,
	S~Zavatarelli, and D.~{}S. Zvezhinskij.
	\newblock {Proposed antimatter gravity measurement with an antihydrogen beam}.
	\newblock {\em Nuclear Instruments and Methods in Physics Research B},
	266:351--356, 2008.
	
	\bibitem{sala2015matter}
	Simone Sala, Fabrizio Castelli, Marco Giammarchi, Stefano Siccardi, and Stefano
	Olivares.
	\newblock Matter-wave interferometry: towards antimatter interferometers.
	\newblock {\em Journal of Physics B: Atomic, Molecular and Optical Physics},
	48(19):195002, 2015.
	
	\bibitem{2016PhRvA..94c3625S}
	S.~{Sala}, M.~{Giammarchi}, and S.~{Olivares}.
	\newblock {Asymmetric Talbot-Lau interferometry for inertial sensing}.
	\newblock {\em \pra}, 94(3):033625, September 2016.
	
	\bibitem{borde1989atomic}
	Ch~J Bord{\'e}.
	\newblock Atomic interferometry with internal state labelling.
	\newblock {\em Physics letters A}, 140(1-2):10--12, 1989.
	
	\bibitem{kasevich1991atomic}
	Mark Kasevich and Steven Chu.
	\newblock Atomic interferometry using stimulated raman transitions.
	\newblock {\em Physical review letters}, 67(2):181, 1991.
	
	\bibitem{RevModPhys.81.1051}
	Alexander~D. Cronin, J\"org Schmiedmayer, and David~E. Pritchard.
	\newblock Optics and interferometry with atoms and molecules.
	\newblock {\em Rev. Mod. Phys.}, 81:1051--1129, Jul 2009.
	
	\bibitem{abend2019atom}
	S~Abend, M~Gersemann, C~Schubert, D~Schlippert, EM~Rasel, M~Zimmermann,
	MA~Efremov, A~Roura, FA~Narducci, and WP~Schleich.
	\newblock Atom interferometry and its applications.
	\newblock {\em Foundations of Quantum Theory}, 197:345, 2019.
	
	\bibitem{hornberger2009theory}
	Klaus Hornberger, Stefan Gerlich, Hendrik Ulbricht, Lucia Hackerm{\"u}ller,
	Stefan Nimmrichter, Ilya~V Goldt, Olga Boltalina, and Markus Arndt.
	\newblock Theory and experimental verification of kapitza--dirac--talbot--lau
	interferometry.
	\newblock {\em New Journal of Physics}, 11(4):043032, 2009.
	
	\bibitem{brezger2002matter}
	Bj{\"o}rn Brezger, Lucia Hackerm{\"u}ller, Stefan Uttenthaler, Julia
	Petschinka, Markus Arndt, and Anton Zeilinger.
	\newblock Matter-wave interferometer for large molecules.
	\newblock {\em Physical review letters}, 88(10):100404, 2002.
	
	\bibitem{gerlich2007kapitza}
	Stefan Gerlich, Lucia Hackerm{\"u}ller, Klaus Hornberger, Alexander Stibor,
	Hendrik Ulbricht, Michael Gring, Fabienne Goldfarb, Tim Savas, Marcel
	M{\"u}ri, Marcel Mayor, et~al.
	\newblock A kapitza--dirac--talbot--lau interferometer for highly polarizable
	molecules.
	\newblock {\em Nature Physics}, 3(10):711, 2007.
	
	\bibitem{mcguirk2000large}
	JM~McGuirk, MJ~Snadden, and MA~Kasevich.
	\newblock Large area light-pulse atom interferometry.
	\newblock {\em Physical review letters}, 85(21):4498, 2000.
	
	\bibitem{2011PhRvL.107m0403C}
	S.-W. {Chiow}, T.~{Kovachy}, H.-C. {Chien}, and M.~A. {Kasevich}.
	\newblock {102${\hbar}$k Large Area Atom Interferometers}.
	\newblock {\em Physical Review Letters}, 107(13):130403, September 2011.
	
	\bibitem{muller2008atom}
	Holger M{\"u}ller, Sheng-wey Chiow, Quan Long, Sven Herrmann, and Steven Chu.
	\newblock Atom interferometry with up to 24-photon-momentum-transfer beam
	splitters.
	\newblock {\em Physical review letters}, 100(18):180405, 2008.
	
	\bibitem{2009PhRvL.102x0403M}
	H.~{M{\"u}ller}, S.-W. {Chiow}, S.~{Herrmann}, and S.~{Chu}.
	\newblock {Atom Interferometers with Scalable Enclosed Area}.
	\newblock {\em Physical Review Letters}, 102(24):240403, June 2009.
	
	\bibitem{clade2009large}
	Pierre Clad{\'e}, Sa{\"\i}da Guellati-Kh{\'e}lifa, Fran{\c{c}}ois Nez, and
	Fran{\c{c}}ois Biraben.
	\newblock Large momentum beam splitter using bloch oscillations.
	\newblock {\em Physical review letters}, 102(24):240402, 2009.
	
	\bibitem{2015PhRvL.115j3001K}
	K.~{Kotru}, D.~L. {Butts}, J.~M. {Kinast}, and R.~E. {Stoner}.
	\newblock {Large-Area Atom Interferometry with Frequency-Swept Raman Adiabatic
		Passage}.
	\newblock {\em Physical Review Letters}, 115(10):103001, September 2015.
	
	\bibitem{machluf2013coherent}
	Shimon Machluf, Yonathan Japha, and Ron Folman.
	\newblock Coherent stern--gerlach momentum splitting on an atom chip.
	\newblock {\em Nature communications}, 4, 2013.
	
	\bibitem{2014EL....10563001M}
	G.~D. {McDonald}, C.~C.~N. {Kuhn}, S.~{Bennetts}, J.~E. {Debs}, K.~S.
	{Hardman}, J.~D. {Close}, and N.~P. {Robins}.
	\newblock {A faster scaling in acceleration-sensitive atom interferometers}.
	\newblock {\em EPL (Europhysics Letters)}, 105:63001, March 2014.
	
	\bibitem{margalit2018realization}
	Yair Margalit, Zhifan Zhou, Or~Dobkowski, Yonathan Japha, Daniel Rohrlich,
	Samuel Moukouri, and Ron Folman.
	\newblock Realization of a complete stern-gerlach interferometer.
	\newblock {\em arXiv preprint arXiv:1801.02708}, 2018.
	
	\bibitem{amit2019t}
	Omer Amit, Yair Margalit, Or~Dobkowski, Zhifan Zhou, Yonathan Japha, Matthias
	Zimmermann, Maxim~A Efremov, Frank~A Narducci, Ernst~M Rasel, Wolfgang~P
	Schleich, et~al.
	\newblock T 3 stern-gerlach matter-wave interferometer.
	\newblock {\em Physical review letters}, 123(8):083601, 2019.
	
	\bibitem{Zimmermann2017}
	M.~Zimmermann, M.~A. Efremov, A.~Roura, W.~P. Schleich, S.~A. DeSavage, J.~P.
	Davis, A.~Srinivasan, F.~A. Narducci, S.~A. Werner, and E.~M. Rasel.
	\newblock T 3-interferometer for atoms.
	\newblock {\em Applied Physics B}, 123(4):102, Mar 2017.
	
	\bibitem{margalit2019analysis}
	Yair Margalit, Zhifan Zhou, Shimon Machluf, Yonathan Japha, Samuel Moukouri,
	and Ron Folman.
	\newblock Analysis of a high-stability stern--gerlach spatial fringe
	interferometer.
	\newblock {\em New Journal of Physics}, 21(7):073040, 2019.
	
	\bibitem{1988OptCo..65..419L}
	E.~P. {Liang} and C.~D. {Dermer}.
	\newblock {Laser cooling of positronium}.
	\newblock {\em Optics Communications}, 65:419--424, March 1988.
	
	\bibitem{1993PhRvL..70.2257S}
	I.~{}D. Setija, H.~{}G.~{}C. Werij, O.~{}J. Luiten, M.~{}W. Reynolds, T.~{}W.
	Hijmans, and J.~{}T.~{}M. Walraven.
	\newblock {Optical cooling of atomic hydrogen in a magnetic trap}.
	\newblock {\em Physical Review Letters}, 70:2257--2260, 1993.
	
	\bibitem{stockton2011absolute}
	JK~Stockton, K~Takase, and MA~Kasevich.
	\newblock Absolute geodetic rotation measurement using atom interferometry.
	\newblock {\em Physical review letters}, 107(13):133001, 2011.
	
	\bibitem{dutta2016continuous}
	I~Dutta, D~Savoie, B~Fang, B~Venon, CL~Garrido Alzar, R~Geiger, and
	A~Landragin.
	\newblock Continuous cold-atom inertial sensor with 1 nrad/sec rotation
	stability.
	\newblock {\em Physical review letters}, 116(18):183003, 2016.
	
	\bibitem{peters2001high}
	Achim Peters, Keng~Yeow Chung, and Steven Chu.
	\newblock High-precision gravity measurements using atom interferometry.
	\newblock {\em Metrologia}, 38(1):25, 2001.
	
	\bibitem{2015PhRvA..91c3629B}
	G.~W. {Biedermann}, X.~{Wu}, L.~{Deslauriers}, S.~{Roy}, C.~{Mahadeswaraswamy},
	and M.~A. {Kasevich}.
	\newblock {Testing gravity with cold-atom interferometers}.
	\newblock {\em \pra}, 91(3):033629, March 2015.
	
	\bibitem{amole2013description}
	C~Amole, MD~Ashkezari, M~Baquero-Ruiz, W~Bertsche, E~Butler, A~Capra, CL~Cesar,
	M~Charlton, S~Eriksson, J~Fajans, et~al.
	\newblock Description and first application of a new technique to measure the
	gravitational mass of antihydrogen.
	\newblock {\em Nature communications}, 4:ncomms2787, 2013.
	
	\bibitem{cui2018time}
	Jiafeng Cui, Yaoyao Xu, Lele Chen, Kun Qi, Minkang Zhou, Xiaochun Duan, and
	Zhongkun Hu.
	\newblock Time base evaluation for atom gravimeters.
	\newblock {\em Review of Scientific Instruments}, 89(8):083104, 2018.
	
	\bibitem{karcher2018improving}
	Romain Karcher, Almazbek Imanaliev, S{\'e}bastien Merlet, and F~Pereira
	Dos~Santos.
	\newblock Improving the accuracy of atom interferometers with ultracold
	sources.
	\newblock {\em New Journal of Physics}, 20(11):113041, 2018.
	
	\bibitem{kritsotakis2018optimal}
	Michail Kritsotakis, Stuart~S Szigeti, Jacob~A Dunningham, and Simon~A Haine.
	\newblock Optimal matter-wave gravimetry.
	\newblock {\em Physical Review A}, 98(2):023629, 2018.
	
	\bibitem{bongs2019taking}
	Kai Bongs, Michael Holynski, Jamie Vovrosh, Philippe Bouyer, Gabriel Condon,
	Ernst Rasel, Christian Schubert, Wolfgang~P Schleich, and Albert Roura.
	\newblock Taking atom interferometric quantum sensors from the laboratory to
	real-world applications.
	\newblock {\em Nature Reviews Physics}, pages 1--9, 2019.
	
	\bibitem{fu2019new}
	Zhijie Fu, Bin Wu, Bing Cheng, Yin Zhou, Kanxing Weng, Dong Zhu, Zhaoying Wang,
	and Qiang Lin.
	\newblock A new type of compact gravimeter for long-term absolute gravity
	monitoring.
	\newblock {\em Metrologia}, 2019.
	
	\bibitem{2013AnP...525..493Y}
	Y.~{Yamazaki} and S.~{Ulmer}.
	\newblock {CPT symmetry tests with cold $\bar p$ and antihydrogen}.
	\newblock {\em Annalen der Physik}, 525:493--504, July 2013.
	
	\bibitem{2013arXiv1304.3721H}
	M.~{Hori} and J.~{Walz}.
	\newblock {Physics at CERN's Antiproton Decelerator}.
	\newblock {\em ArXiv e-prints}, April 2013.
	
	\bibitem{2014IJMPS..3060258K}
	K.~{Kirch} and K.~S. {Khaw}.
	\newblock {Testing antimatter gravity with muonium}.
	\newblock In {\em International Journal of Modern Physics Conference Series},
	volume~30 of {\em International Journal of Modern Physics Conference Series},
	page 1460258, May 2014.
	
	\bibitem{2015PhRvD..92e6002K}
	V.~A. {Kosteleck{\'y}} and A.~J. {Vargas}.
	\newblock {Lorentz and C P T tests with hydrogen, antihydrogen, and related
		systems}.
	\newblock {\em \prd}, 92(5):056002, September 2015.
	
	\bibitem{2017arXiv171001833S}
	M.~S. {Safronova}, D.~{Budker}, D.~{DeMille}, D.~F.~J. {Kimball},
	A.~{Derevianko}, and C.~W. {Clark}.
	\newblock {Search for New Physics with Atoms and Molecules}.
	\newblock {\em ArXiv e-prints}, October 2017.
	
	\bibitem{poggiani1993possible}
	R~Poggiani.
	\newblock A possible gravity measurement with antihydrogen.
	\newblock {\em Hyperfine Interactions}, 76(1):371--377, 1993.
	
	\bibitem{poggiani1997measurement}
	Rosa Poggiani.
	\newblock Measurement of the gravitational acceleration of antihydrogen.
	\newblock {\em Hyperfine interactions}, 109(1):367--372, 1997.
	
	\bibitem{Mills2002}
	A.~P. {Mills} and M.~{Leventhal}.
	\newblock Can we measure the gravitational free fall of cold rydberg state
	positronium?
	\newblock {\em Nuclear Instruments and Methods in Physics Research B},
	192:102--106, May 2002.
	
	\bibitem{walz2004proposal}
	Jochen Walz and Theodor~W H{\"a}nsch.
	\newblock A proposal to measure antimatter gravity using ultracold antihydrogen
	atoms.
	\newblock {\em General Relativity and Gravitation}, 36(3):561--570, 2004.
	
	\bibitem{perez2005new}
	P~Perez and A~Rosowsky.
	\newblock A new path toward gravity experiments with antihydrogen.
	\newblock {\em Nuclear Instruments and Methods in Physics Research Section A:
		Accelerators, Spectrometers, Detectors and Associated Equipment},
	545(1):20--30, 2005.
	
	\bibitem{2014IJMPS..3060259C}
	D.~B. {Cassidy} and S.~D. {Hogan}.
	\newblock {Atom control and gravity measurements using Rydberg positronium}.
	\newblock {\em International Journal of Modern Physics Conference Series},
	30:1460259, May 2014.
	
	\bibitem{perez2015gbar}
	Patrice Perez, D~Banerjee, F~Biraben, D~Brook-Roberge, M~Charlton, P~Clad{\'e},
	P~Comini, P~Crivelli, O~Dalkarov, P~Debu, et~al.
	\newblock The gbar antimatter gravity experiment.
	\newblock {\em Hyperfine Interactions}, 233(1-3):21--27, 2015.
	
	\bibitem{2014IJMPS..3060257C}
	P.~{Crivelli}, D.~A. {Cooke}, and S.~{Friedreich}.
	\newblock {Experimental considerations for testing antimatter antigravity using
		positronium 1S-2S spectroscopy}.
	\newblock {\em International Journal of Modern Physics Conference Series},
	30:1460257, May 2014.
	
	\bibitem{1995leap.conf..569P}
	T.~J. {Phillips}.
	\newblock {A Technique for Directly Measuring the Gravitational Acceleration of
		Antihydrogen}.
	\newblock In G.~{Kernel}, P.~{Krian}, and M.~{Miku}, editors, {\em Low Energy
		Antiproton Physics}, page 569, 1995.
	
	\bibitem{1996HyInt.100..163P}
	T.~J. {Phillips}.
	\newblock {Measuring the gravitational acceleration of antimatter with an
		antihydrogen interferometer}.
	\newblock {\em Hyperfine Interactions}, 100:163--172, December 1996.
	
	\bibitem{phillips1997antimatter}
	Thomas~J Phillips.
	\newblock Antimatter gravity studies with interferometry.
	\newblock {\em Hyperfine interactions}, 109(1):357--365, 1997.
	
	\bibitem{chang1975space}
	Byung~Jin Chang, R~Alferness, and Emmett~N Leith.
	\newblock Space-invariant achromatic grating interferometers: theory.
	\newblock {\em Applied optics}, 14(7):1592--1600, 1975.
	
	\bibitem{2005JPhB...38.1765H}
	M.~{}W. Horbatsch, M~Horbatsch, and E.~{}A. Hessels.
	\newblock {A universal formula for the accurate calculation of hydrogenic
		lifetimes}.
	\newblock {\em Journal of Physics B Atomic Molecular Physics}, 38:1765--1771,
	2005.
	
	\bibitem{2016PhyS...91e3006B}
	B.~{Barrett}, A.~{Bertoldi}, and P.~{Bouyer}.
	\newblock {Inertial quantum sensors using light and matter}.
	\newblock {\em \physscr}, 91(5):053006, May 2016.
	
	\bibitem{mcdonald2014faster}
	Gordon~D McDonald, Carlos~CN Kuhn, Shayne Bennetts, John~E Debs, Kyle~S
	Hardman, John~D Close, and Nicholas~P Robins.
	\newblock A faster scaling in acceleration-sensitive atom interferometers.
	\newblock {\em EPL (Europhysics Letters)}, 105(6):63001, 2014.
	
	\bibitem{zimmermann2019representation}
	Matthias Zimmermann, Maxim~A Efremov, Wolfgang Zeller, Wolfgang~P Schleich,
	Jon~P Davis, and Frank~A Narducci.
	\newblock Representation-free description of atom interferometers in
	time-dependent linear potentials.
	\newblock {\em New Journal of Physics}, 21(7):073031, 2019.
	
	\bibitem{Orloff2008}
	Jon Orloff.
	\newblock {\em Handbook of Charged Particle Optics}.
	\newblock CRC Press, 2008.
	
	\bibitem{gallagher1994}
	Thomas~F Gallagher.
	\newblock {\em {Rydberg Atoms}}.
	\newblock Cambridge University Press, Cambridge, 1994.
	
	\bibitem{friedrich1989hydrogen}
	Harald Friedrich and Hieter Wintgen.
	\newblock The hydrogen atom in a uniform magnetic field - an example of chaos.
	\newblock {\em Physics Reports}, 183(2):37--79, 1989.
	
	\bibitem{pinard1990atoms}
	J~Pinard.
	\newblock Atoms in static electric and magnetic fields: The experimental
	aspect.
	\newblock In {\em Atoms in Strong Fields}, pages 17--42. Springer, 1990.
	
	\bibitem{lisitsa1987new}
	Valerii~S Lisitsa.
	\newblock New results on the stark and zeeman effects in the hydrogen atom.
	\newblock {\em Soviet Physics Uspekhi}, 30(11):927, 1987.
	
	\bibitem{Bartsch2006}
	Thomas Bartsch and Turgay Uzer.
	\newblock {\em Rydberg Atoms in Strong Static Fields}, pages 247--252.
	\newblock Springer New York, New York, NY, 2006.
	
	\bibitem{PhysRevA.57.1149}
	J\"{o}rg Main, Michael Schwacke, and G\"{u}nter Wunner.
	\newblock {Hydrogen atom in combined electric and magnetic fields with
		arbitrary mutual orientations}.
	\newblock {\em Phys. Rev. A}, 57(2):1149--1157, 1998.
	
	\bibitem{solovev1983second}
	EA~Solovev.
	\newblock Second order perturbation theory for the hydrogen atom in crossed
	electric and magnetic fields.
	\newblock {\em Zhurnal Eksperimental'noi i Teoreticheskoi Fiziki}, 85:109--114,
	1983.
	
	\bibitem{demkov1970energy}
	Yu~N Demkov, BS~Monozon, and VN~OSTROVSKil.
	\newblock Energy levels of a hydrogen atom in crossed electric and magnetic
	fields.
	\newblock {\em Soviet Physics JETP}, 30(4), 1970.
	
	\bibitem{heupel2002hydrogen}
	T~Heupel, M~Mei, M~Niering, B~Gross, M~Weitz, TW~H{\"a}nsch, and Ch~J
	Bord{\'e}.
	\newblock Hydrogen atom interferometer with short light pulses.
	\newblock {\em EPL (Europhysics Letters)}, 57(2):158, 2002.
	
	\bibitem{palmer2019electric}
	JE~Palmer and SD~Hogan.
	\newblock Electric rydberg-atom interferometry.
	\newblock {\em Physical Review Letters}, 122(25):250404, 2019.
	
	\bibitem{palmer2019matter}
	JE~Palmer and SD~Hogan.
	\newblock Matter-wave interferometry with atoms in high rydberg states.
	\newblock {\em Molecular Physics}, pages 1--12, 2019.
	
	\bibitem{1978PhRvA..18.1853S}
	H.~J. {Silverstone}.
	\newblock {Perturbation theory of the Stark effect in hydrogen to arbitrarily
		high order}.
	\newblock {\em \pra}, 18:1853--1864, November 1978.
	
	\bibitem{PhysRevA.47.1209}
	C.~Bordas and H.~Helm.
	\newblock Electric-field ionization of rydberg states of ${\mathrm{h}}_{3}$.
	\newblock {\em Phys. Rev. A}, 47:1209--1219, Feb 1993.
	
	\bibitem{PhysRevA.87.063423}
	S.~D. Hogan.
	\newblock Calculated photoexcitation spectra of positronium rydberg states.
	\newblock {\em Phys. Rev. A}, 87:063423, Jun 2013.
	
	\bibitem{ovsiannikov1998diamagnetic}
	VD~Ovsiannikov and SV~Goossev.
	\newblock Diamagnetic shift and splitting of rydberg levels in atoms.
	\newblock {\em Physica Scripta}, 57(4):506, 1998.
	
	\bibitem{borde2014atom}
	Christian~J Bord{\'e}.
	\newblock Atom interferometry using internal excitation: Foundations and recent
	theory.
	\newblock {\em International School of Physics Enrico Fermi--COURSE CLXXXVIII
		Atom Interferometry}, pages 143--170, 2014.
	
	\bibitem{borde2001theoretical}
	Christian~J Bord{\'e}.
	\newblock Theoretical tools for atom optics and interferometry.
	\newblock {\em Comptes Rendus de l'Academie des Sciences-Series IV-Physics},
	2(3):509--530, 2001.
	
	\bibitem{antoine2003quantum}
	Ch~Antoine and Ch~J Bord{\'e}.
	\newblock Quantum theory of atomic clocks and gravito-inertial sensors: an
	update.
	\newblock {\em Journal of Optics B: Quantum and Semiclassical Optics},
	5(2):S199, 2003.
	
	\bibitem{dubetsky2006atom}
	B~Dubetsky and MA~Kasevich.
	\newblock Atom interferometer as a selective sensor of rotation or gravity.
	\newblock {\em Physical Review A}, 74(2):023615, 2006.
	
	\bibitem{2014NJPh...16l3012R}
	A.~{Roura}, W.~{Zeller}, and W.~P. {Schleich}.
	\newblock {Overcoming loss of contrast in atom interferometry due to gravity
		gradients}.
	\newblock {\em New Journal of Physics}, 16(12):123012, December 2014.
	
	\bibitem{kleinert2015representation}
	Stephan Kleinert, Endre Kajari, Albert Roura, and Wolfgang~P Schleich.
	\newblock Representation-free description of light-pulse atom interferometry
	including non-inertial effects.
	\newblock {\em Physics Reports}, 605:1--50, 2015.
	
	\bibitem{roura2017circumventing}
	Albert Roura.
	\newblock Circumventing heisenberg's uncertainty principle in atom
	interferometry tests of the equivalence principle.
	\newblock {\em Physical Review Letters}, 118(16):160401, 2017.
	
	\bibitem{chaneliere2018phase}
	Thierry Chaneli{\`e}re, Daniel Comparat, and Hans Lignier.
	\newblock Phase-space-density limitation in laser cooling without spontaneous
	emission.
	\newblock {\em Physical Review A}, 98(6):063432, 2018.
	
	\bibitem{kennard1927quantenmechanik}
	Earle~H Kennard.
	\newblock Zur quantenmechanik einfacher bewegungstypen.
	\newblock {\em Zeitschrift f{\"u}r Physik A Hadrons and Nuclei},
	44(4):326--352, 1927.
	
	\bibitem{van1928correspondence}
	John~H Van~Vleck.
	\newblock The correspondence principle in the statistical interpretation of
	quantum mechanics.
	\newblock {\em Proceedings of the National Academy of Sciences},
	14(2):178--188, 1928.
	
	\bibitem{morette1951definition}
	C{\'e}cile Morette.
	\newblock On the definition and approximation of feynman's path integrals.
	\newblock {\em Physical Review}, 81(5):848, 1951.
	
	\bibitem{case2008wigner}
	William~B Case.
	\newblock Wigner functions and weyl transforms for pedestrians.
	\newblock {\em American Journal of Physics}, 76(10):937--946, 2008.
	
	\bibitem{storey1994feynman}
	Pippa Storey and Claude Cohen-Tannoudji.
	\newblock The feynman path integral approach to atomic interferometry. a
	tutorial.
	\newblock {\em Journal de Physique II}, 4(11):1999--2027, 1994.
	
	\bibitem{bongs2006high}
	Kai Bongs, R~Launay, and Mark~A Kasevich.
	\newblock High-order inertial phase shifts for time-domain atom
	interferometers.
	\newblock {\em Applied Physics B}, 84(4):599--602, 2006.
	
	\bibitem{bertoldi2019phase}
	A~Bertoldi, F~Minardi, and M~Prevedelli.
	\newblock Phase shift in atom interferometers: Corrections for nonquadratic
	potentials and finite-duration laser pulses.
	\newblock {\em Physical Review A}, 99(3):033619, 2019.
	
\end{thebibliography}

\end{document}